\documentclass[twocolumn,showpacs,showkeys,amsmath,amssymb]{revtex4}
\usepackage[dvips]{graphicx}
\usepackage{amsthm}
\usepackage{amscd}
\usepackage{amstext}
\usepackage{amsfonts}
\usepackage{hyperref}
\newcommand{\p}{\partial}

\newcommand{\dd}{{\rm d}}

\begin{document}
%u
%\preprint{}

\title[Spacetime connectedness and Lorentz force equation]{Comment on ``The {A}vez-{S}eifert theorem for the relativistic {L}orentz force equation" and other related works} % Force line breaks with \\

\author{E. Minguzzi}
 \affiliation{Departamento de Matem\'aticas,
Universidad de Salamanca, \\ Plaza de la Merced 1-4, E-37008
Salamanca, Spain \\ and INFN, Piazza dei Caprettari 70, I-00186
Roma, Italy
\\ minguzzi@usal.es }%Lines break automatically or can be forced with \\
%\eads{}
%\date{May 4, 2004}

\begin{abstract}
There exist several approaches that investigate the connectedness
of spacetime events through solutions of the  Lorentz force
equation. These approaches separate into three categories, that
consider different equations. We clarify the  physical meaning of
each equation showing that only one method is based on the Lorentz
force equation. The other two approaches lead respectively to a
less restrictive equation that defines an electromagnetic flow on
the cotangent fiber bundle, or to an unphysical constraint between
charge-to-mass ratio and proper length of the solution. We outline
the physical meaning of each approach studying the  variational
formulations and clarifying the results obtained in the explored
directions.
\end{abstract}

%Use showkeys class option if keyword
                              %display desired
\maketitle

\newpage

\section{Introduction}
Recently,  a work by E. Caponio and A. Masiello,  in some articles
already cited as \cite{caponio01}, and announced in
\cite{masiello02,caponio04b} has appeared, with minor
modifications, in the pages of the Journal of Mathematical Physics
\cite{caponio04d}. The article deals with the problem of
connectedness of a globally hyperbolic spacetime through solutions
of the Lorentz force equation. This problem attracted some
attention in the last years since a positive answer, i.e. the
proof of the statement {\em in a globally hyperbolic spacetime,
given two chronologically related events, there is a solution of
the Lorentz force equation passing through them}, would generalize
to charged particles a well-known theorem by Avez and Seifert
\cite[theorem 3.18]{beem96}, \cite[proposition 14.19]{oneill83},
\cite[proposition 6.7.1]{hawking73}. Caponio and Masiello's work,
now published, had the merit of introducing a Kaluza-Klein
approach to deal with
 this geometrical problem.
%In this framework the solutions of the Lorentz force equation as
%projections of geodesics on a higher dimensional bundle.
Unfortunately, it shares a problem, with some other works on the
same subject,  concerning the physical interpretation of the
results obtained. This problem arises since while in the Lorentz
force equation the 4-velocity should be {\em a priori} normalized,
usually with conditions as $u^{\alpha}u_{\alpha}=1$ or $p^{\alpha}
p_{\alpha}=m^2$, in Caponio and Masiello's article  this condition
is dropped leading to different physical and mathematical problems
(see subsections A and C of the bibliography). Indeed, dropping
the normalization condition the space of solutions becomes
infinitely larger than in the usual Lorentz force equation and it
becomes easier to prove the connectedness through solutions of the
modified equation. While these  modified problems have a
mathematical  interest in their own right and can be studied as
such \cite{antonacci00,bartolo01}, results on them have been
sometimes improperly ascribed
 as results on  the more restrictive Lorentz force equation
\cite{bartolo00,bartolo03,caponio02,caponio02b,caponio04c,caponio04d}
leading to
%Since, despite this modification, the results in this direction
%appear in those works, with an incorrect terminology, as results
%on the Lorentz force equation, we are now in
a condition where, apparently, the same claims and theorems are
published twice in different journals (compare
\cite{caponio04d,minguzzi03b}). In fact some of those results are
related to the Lorentz force equation (subsection B of the
bibliography) while others are related to less restrictive
equations or different constraints (subsections A and C of the
bibliography). In this comment we try to clarify this complex
situation putting into perspective the results obtained in the
different directions.

\section{Lorentz force equation (LFE)}
The Lorentz force equation describes the motion of a charged
particle over spacetime in presence of an electromagnetic field.
It describes the motion of ideal spinless, pointlike particles
when the radiative corrections and all the quantum effects can be
ignored.

 Consider a spacetime $M$, that is a time-oriented $n$-dimensional
Lorentzian manifold endowed with a metric $g$ with signature $(+
-,\cdots,-)$, and an electromagnetic field $F$ on $M$, that is, a
skew symmetric closed 2-form.  Let $c=1$. A point particle of rest
mass $m>0$ and electric charge $q\in \mathbb{R}$, moving under the
action of the field $F$, has a worldline which satisfies the {\em
Lorentz force equation} (LFE) (cf. \cite{jackson75,misner73})
\begin{equation} \label{lorentz}
 D_s\!\left(\frac{\dd x}{\dd s}\right)=\frac{q}{m}\hat F(x)\left[\frac{\dd x}{\dd s}\right].
\end{equation}
Here $x=x(s)$ is the worldline of the particle parameterized with
respect to the proper length, $\frac{\dd x}{\dd s}$ is the
$n$-velocity, $D_s\!\left(\frac{\dd x}{\dd s}\right)$ is the
covariant derivative of $\frac{\dd x}{\dd s}$ along $x(s)$
associated to the Levi-Civita connection of $g$, and $\hat
F(x)[\cdot]$ is the linear map  on $T_x M$ obtained raising the
left-hand index of $F$. It is understood that the Lorentz force
equation is determined only once  the parameter $q/m$ has been
given.  The solutions of the Lorentz force equation are
future-oriented timelike curves parametrized with respect to
proper length. They are interpreted as trajectories of the
particle with the given ratio $q/m$ on spacetime. These solution
may be regarded as $C^{2}$ mappings $x$ from an interval of the
real line to $M$. Given such mapping $x(\lambda)$ all the others
obtained from this one by an orientation-preserving
reparametrization, $\lambda=f(\lambda')$, $f'>0$, are regarded as
physically equivalent. In other words the parametrization assures
only the differentiability of the curve, what matters physically
is the image of the application, that is, the trajectory. Note,
indeed, that any mathematical model willing to describe the motion
of a particle should face with what can be actually observed. The
principal observable in the motion of a charge is its trajectory
on spacetime. Indeed, knowing the spacetime under consideration
and the electromagnetic field, from the trajectory one usually
recovers the charge-to-mass ratio (using (\ref{lorentz})) and the
proper time of the particle (integrating the line element over the
trajectory). The proper time so obtained can sometimes be compared
with the one measured directly from the decay of a charged
particle previously created, showing that the previous method is
consistent with observations.
%We stress that the parameter
%$\lambda$ to be used for the mathematical representation of the
%curve has in general no meaning because of the freedom in choosing
%the parametrization. The only parameter along the curve that can
%be observed is proper time and derives from the trajectory alone.

A fundamental aspect of Eq. (\ref{lorentz}) is that it contains a
second derivative on the left-hand side and a first derivative on
the right-hand side. This implies that the curve $x$ should be
parametrized with respect to proper time $s$ in order to infer
what is the corresponding charge-to-mass ratio.

Note that given a timelike solution $x(s)$ of (\ref{lorentz}) the
charge-to-mass ratio $q/m$ is uniquely determined unless $x(s)$
satisfies
\begin{equation}\label{frb}
\hat F(x)\left[\frac{\dd x}{\dd s}\right]=0,
\end{equation}
and, then
\begin{equation}\label{fra}
D_s \left(\frac{\dd x}{\dd s}\right)=0.
\end{equation}
%
%\begin{eqnarray}
% D_s\!\left(\frac{\dd x}{\dd s}\right)&=&0, \label{fra} \\
% \hat F(x)\left[\frac{\dd x}{\dd s}\right]&=&0. \label{frb}
%\end{eqnarray}
That is, $x(s)$ is a geodesic of $M$ whose tangent vectors stay in
the $\textrm{ker}\hat F$. If we restrict our analysis to solutions
of (\ref{lorentz}) connecting two chronologically related events,
$x_{0} \ll x_{1}$ this situation occurs in very special cases and
physically is not of primary interest. In fact given the geodesics
connecting the two events, suppose that (\ref{frb}) is satisfied
by one of them, we see that this condition it is spoiled under a
suitable small perturbation $\delta \hat F$ i.e. condition
(\ref{frb}), if regarded as a condition on the electromagnetic
field, is unstable. However, we shall include this case in our
study. Let $X$ be the set  of $C^{2}$ timelike curves  that
satisfy (\ref{lorentz}) for a certain $q/m$ (each one with their
own $q/m$). We define a charge-to-mass ratio function
$\frac{\textrm{q}}{\textrm{m}}: X \to \mathbb{R} \cup \{R\}$ where
$R$ here is just a symbol in the following way: if the curve $x
\in X$ does not satisfy  (\ref{frb}), then we define (we use the
roman letters for the function)
$\frac{\textrm{q}}{\textrm{m}}(x)={q}/{m}$ where $q/m$ is derived
from (\ref{lorentz}); if the curve $x$ satisfies
 (\ref{frb}),  then we define
$\frac{\textrm{q}}{\textrm{m}}(x)=R$.  In this way the function
$\frac{\textrm{q}}{\textrm{m}}(x)$ becomes an observable. The case
$\frac{\textrm{q}}{\textrm{m}}(x)=R$ happens when from the
observation of $x$ the observer can not infer the {\em real} (in
mathematical sense) value of the charge-to-mass ratio of the
particle. If $\frac{\textrm{q}}{\textrm{m}}(x)=R$, then $x$ solves
(\ref{lorentz}) independently of the value of $q/m$.

Finally, let $F$ be an exact electromagnetic field $F=\dd \omega$.
Recall that the Lorentz force equation is satisfied by any
timelike stationary point of the charged-particle action
\cite[section 16]{landau62} (we write for short $\dd s =
\sqrt{g_{\mu \nu}\frac{\dd x^{\mu}}{\dd \lambda} \frac{\dd
x^{\nu}}{\dd \lambda}} \dd \lambda$)
\begin{equation}\label{fun}
I_{x_0,x_1}[\gamma]=\int _{\gamma} (\dd s+\frac{q}{m}  \omega),
\end{equation}
defined on the set of $C^{1}$ causal curves connecting $x_{0}$ and
$x_{1}$. Thus, in order to prove the existence of connecting
solutions of the  Lorentz force equation one can look for timelike
stationary points of this functional.

\section{The electromagnetic flow equation (EFE)}
In order to describe the motion of a charged particle we have to
determine the set of trajectories of its motion on spacetime. It
is clear from the previous equations that the set of solutions
will differ only for particles having a different charge-to-mass
ratio and that at least looking at the motion of the particle,
neither the mass nor the charge are separately observable
%\footnote{Of course, only as long as the physics of the Lorentz
%force equation is concerned. At the variational level one should
%add more terms to the charged-particle action in order to make
%both charge and mass observables. This, however, would lead us to
%a different mathematical problem.}
\cite{endnote30}. Consider the equation
\begin{equation} \label{lorentz2}
 D_{\lambda} \left(\frac{\dd x}{\dd
\lambda}\right)=Q\hat F(x)\left[\frac{\dd x}{\dd \lambda}\right],
\end{equation}
with $\lambda$ a dimensional parameter (in some works
\cite{bartolo99,bartolo03,caponio02,caponio04d,mirenghi02} the
letter $s$ is used in place of $\lambda$ but this does not mean
that in those works $s$ is the proper time). Its dimension is
chosen in such a way that $Q$ has the dimension of a charge. The
solutions of this equation are mappings $x:(\lambda_{0},
\lambda_{1}) \to M$. This equation is referred, in
\cite{bartolo99,bartolo00,bartolo03,caponio02,caponio02b,caponio04,caponio04c,caponio04d,mirenghi02},
as the {\em Lorentz force
%\cite{guillemin84,caponio01,caponio02,caponio04,caponio04c,piccione04}
 equation} of a particle of charge $Q$ (sometimes normalized). We wish
to show that it is inappropriate to call it {\em Lorentz force
equation}, as this terminology could lead to confusion both from
the physical and mathematical side. In order to understand what
represents a solution of Eq. (\ref{lorentz2}) let us simply take a
solution $x(\lambda)$, parametrize it with proper time and
substitute it on (\ref{lorentz}). First of all, Eq.
(\ref{lorentz2}) implies that
\begin{equation}
g_{\mu \nu}\frac{\dd x^{\mu}}{\dd \lambda} \frac{\dd x^{\nu}}{\dd
\lambda}=(\frac{\dd s}{\dd \lambda})^{2}=C^{2}
\end{equation}
that is the parameter $\lambda$ is related to proper length by the
relation $C \dd \lambda=\dd s$ where $C$ is a constant. We
restrict to the timelike solutions, the only ones that may receive
 the interpretation of massive particles, and therefore restrict
our attention to the case $C \in \mathbb{R}$, $C>0$. Replacing the
previous equation in (\ref{lorentz}) we find
\begin{equation}
 D_s\!\left(\frac{\dd x}{\dd s}\right)=\frac{Q}{C}\hat F(x)\left[\frac{\dd x}{\dd s}\right].
\end{equation}
and hence the charge-to-mass ratio is $q/m=Q/C$. All the problems
with Eq. (\ref{lorentz2}) arise from the fact that the constant
$C$ is not fixed a priori leading to solutions of arbitrary
charge-to-mass ratios. Solutions to this equation cannot be
associated to trajectories of a given particle.

\subsubsection{Problems involving the ``charge" $Q$} If we
characterize Eq. (\ref{lorentz2}) with a parameter $Q$ it should
at least have some physical meaning in the solutions of
that equation.  %The charge is not observable so, if this is the
%meaning of $Q$, it seems to be difficult to find a trace of this
%coefficient in the solutions. No statement regarding the charge
%alone could never makes sense in this context.
Consider the equation (\ref{lorentz2}) with $Q$ replaced with
$Q'\ne Q$, $\textrm{sgn}(Q)=\textrm{sgn}(Q')$. If $x(\lambda)$ is
a solution of (\ref{lorentz2}), $x(\frac{Q}{Q'}\lambda)$ is a
solution of (\ref{lorentz2}) with $Q$ replaced with $Q'$. However,
the trajectories are exactly the same so the set of trajectories
solutions of (\ref{lorentz2})  is independent of the value of $Q$.
It depends only on $\textrm{sgn}(Q)$. Thus there is no reason to
call Eq. (\ref{lorentz2}) the Lorentz force equation of charge $Q$
since its solutions when regarded as trajectories (ultimately the
only observable) do not depend from $\vert Q \vert $. It is
preferable to replace $Q \to \epsilon=\textrm{sgn}(Q)$ and change
the dimensionality of $\lambda$ to obtain, given $\epsilon=\pm 1$,
\begin{equation}
D_{\lambda} \left(\frac{\dd x}{\dd \lambda}\right)= \epsilon \hat
F(x)\left[\frac{\dd x}{\dd \lambda}\right].
\end{equation}
We see therefore that Eq. (\ref{lorentz2}) displays a coefficient
that has in fact no clear physical meaning and that does not
select different spaces of solutions although the notation could
suggest the contrary. These observations point out that, in
general, if a mathematical model is designed to describe a
physical situation, as far as possible, only observables should
enter the construction, and in particular the coefficients of the
model should have some physical consequence in order to receive a
physical interpretation. In this case the coefficient $Q$ in
(\ref{lorentz2}) cannot receive an interpretation and should be
better removed as suggested.

\subsubsection{Problems involving the charge-to-mass ratios}
Consider the motion of two whatever particles, with  the same sign
of their charge-to-mass ratios. Their trajectories solve the
Lorentz force equation (\ref{lorentz}) each one with their
respective charge-to-mass ratios. However, both trajectories can
be easily parametrized to give a solution of {\em the same}
equation (\ref{lorentz2}) with $Q$ of the same sign of their
charge-to-mass ratios. It suffices to take as parameter $\dd
\lambda=\frac{ q}{Q m} \dd s$ where $q$ and $m$ are the charge and
the mass of the particle that follows the trajectory considered.
Thus Eq. (\ref{lorentz2}) is unable to distinguish between
solutions with different charge-to-mass ratios. Any trajectory
solution of a Lorentz force equation relative to a certain
charge-to-mass ratio with $\textrm{sgn}(q/m)=\textrm{sgn}(Q)$ is
solution of (\ref{lorentz2}), while the converse is of course not
true.\\
%This difference is crucial when considering the problem of
%connectedness of spacetime events, as the set of solutions of the
%Lorentz force equation is far smaller than that of Eq.
%(\ref{lorentz2}).\\

In order to solve these problems the only way out is to add a
constraint that should be satisfied by every solution. Eq.
(\ref{lorentz2}) should be coupled with the equation
\begin{equation} \label{bla}
g_{\mu \nu}\frac{\dd x^{\mu}}{\dd \lambda} \frac{\dd x^{\nu}}{\dd
\lambda}=(\frac{\dd s}{\dd \lambda})^{2}=m^{2}.
\end{equation}
Note that while any solution of Eq. (\ref{lorentz2}) implies that
$\dd s/\dd \lambda$ is a constant, here we are fixing its value
{\em a priori} and thus are removing many solutions of the
original equation \cite{sachs77}. The system so obtained is
clearly equivalent to the original Lorentz force equation but
should be better avoided since there appear three unobservable
quantities (at least looking at the motion of the particle)  $Q$,
$m$, and $\lambda$ while in the Lorentz force equation all the
coefficients appear in truly observable combinations. In the works
mentioned no a priori constraint is imposed  on the square of the
4-velocity. Some authors
%Unfortunately, in
%the articles mentioned Eq. (\ref{lorentz2}) alone is called
%Lorentz force equation of a charge $Q$ and not in system with
%(\ref{bla}). In those papers \cite{caponio02,caponio04c} the
%authors do not impose any constraint on the square of the
%4-velocity like (\ref{bla}), moreover, sometimes
\cite{caponio04c,caponio04d,bartolo00} refer to \cite{sachs77} as
a source for their terminology while Sachs and Wu define the
Lorentz force equation correctly as a system of (\ref{lorentz2})
with (\ref{bla}) \cite[definitions 3.1.1 and 3.8.1]{sachs77}. Note
that even regarding (\ref{lorentz2}) as an equation of a system it
remains misguiding to call it Lorentz force equation of a charge
$Q$, as in this way one can easily forget that this definition is
correct only inside the system.

\subsection{The electromagnetic flow}
As we said, most problems arise because the set of solutions of
(\ref{lorentz2}) is larger than the one of (\ref{lorentz}). We may
say that the first equation is solved by the trajectory of every
charged particle with a charge-to-mass ratio of the same sign of
$\epsilon$. It is therefore natural that it can be recast in a
form where no coefficient $Q$ appears
\begin{equation} \label{ahh}
D_{\lambda} \left(\frac{\dd x}{\dd \lambda}\right)=\epsilon \hat
F(x)\left[\frac{\dd x}{\dd \lambda}\right].
\end{equation}
This equation can be studied in its own right
%\footnote{Since we
%regard the solution always as {\em future-oriented} trajectories
%we have in fact two flow equations depending on the value of
%$\epsilon$.}
\cite{endnote31}. It needs however a different name. We suggest
{\em electromagnetic flow equation} (EFE) in  analogy with the
equation of the {\em geodesic flow}. Indeed, let us introduce the
quantity $p_{\alpha}=g_{\alpha \beta} \dd x^{\beta}/\dd \lambda$
(we use the letter $p$ since this is a one-form i.e. it lives in
$T^{*}M$; it has not the dimensions of a momentum), then Eq.
(\ref{ahh}) can be rewritten
\begin{eqnarray}
D_{\lambda} \left( p \right)&=& \epsilon \hat
F(x)\left[ p \right] \\
\frac{\dd x^{\alpha}}{ \dd \lambda}&=& p^{\alpha}
\end{eqnarray}
This equation determines a flow in $T^{*}M$. The trajectories of
this flow when projected on $M$ satisfy Eq. (\ref{ahh}). There is,
however, a relevant difference with respect to the geodesic flow.
In fact in that case the trajectories starting from two points in
the same fiber $T^{*}M_{x}$, say $(x, p)$, $(x,p')$, with
$p'=\alpha p$, $\alpha \in \mathbb{R}$ project on the same
trajectory over $M$ while this does not happen for the
electromagnetic flow. For this reason the solutions of (\ref{ahh})
are infinitely more numerous than those of the Lorentz force
equation. This difference is crucial if one tries to prove  the
connectedness of spacetime. \\

%: given an event $x_{0}$ and an event $x_{1}$ in its chronological
%future, determine whether there are solutions of the Lorentz force
%equation that connect them. Clearly the same problem for Eq.
%(\ref{lorentz2}) is easier since a connecting trajectory of Eq.
%(\ref{lorentz}) with $\textrm{sgn}(q/m)=\textrm{sgn}(Q)$ is a
%connecting trajectory of Eq. (\ref{lorentz2}) while the converse
%is not true. Since the literature on both problems has growth in
%the last years, and because of the inappropriate terminology that
%refers to (\ref{lorentz2}), there are results that present the
%same apparent statement while the physical content is really
%different\cite{caponio04d,minguzzi03b}.

%\subsection{Existence results for the Lorentz force equation are multiplicity results\\ for the electromagnetic flow equation}
%Finally we wish to comment an aspect of \cite{piccione04}. In that
%work the authors are interested in multiplicity results for
%connecting trajectories of (\ref{lorentz2}). They briefly comment
%on the usefulness of the Kaluza-Klein approach (the one used in
%\cite{caponio01,caponio03, minguzzi03b}) in approaching this kind
%of questions maintaining that at least in this case that method is
%not really effective. They, however, do not seem to realize that
%the equation considered in \cite{caponio03,minguzzi03b} is not
%(\ref{lorentz2}) but rather (\ref{lorentz}), and that the {\em

\noindent {\bf Remark}.
%Let us consider again the problem of the
%connectedness of spacetime through solutions of the Lorentz force
%equation and through solutions of the electromagnetic flow
%equation. In this section
For instance we show that {\em existence results} for the Lorentz
force equation  (\ref{lorentz}) are in fact {\em multiplicity
results} for the electromagnetic flow equation  (\ref{lorentz2}).

Consider two events $x_{0} \ll x_{1}$ in a globally hyperbolic
spacetime and let $F$ be exact, in \cite{caponio03} it is proved
that there is an interval $U=(-r,r)$ of the real line such that
for each $q/m \in U$ there is a connecting solution of the Lorentz
force equation (\ref{lorentz}). This result was improved in
\cite{minguzzi03b} where it was shown that $r=+\infty$ if there is
no null  geodesic connecting $x_{0}$ and $x_{1}$. We have already
seen that, suitably parametrized, a solution of (\ref{lorentz})
with $\textrm{sgn}(q/m)=\textrm{sgn}(Q)$ becomes solution of
(\ref{lorentz2}), but we can say more: solutions of Lorentz force
equations with different charge-to-mass ratios are distinct unless
a very special case. Indeed, let $x(s)$ be solution of
(\ref{lorentz}) for a charge-to-mass ratio $q/m$ and let $x'(s)$
be a solution of  (\ref{lorentz}) with coefficient $q'/m'\ne q/m$.
Suppose $x'=x$ and subtract the two Lorentz force equations. Since
$\Delta (q/m) \ne 0$ we easily find that $x$ satisfy the system
(\ref{frb}), (\ref{fra}). The existence of such geodesics, as we
already said, happens only if the electromagnetic field satisfies
a very restrictive constraint. For all the other cases we conclude
that there is an infinite degeneracy of connecting solutions of
(\ref{lorentz2}): if $\textrm{sgn}(Q)>0$ (resp.
$\textrm{sgn}(Q)<0$) there is a solution for each $q/m \in (0,r)$
(resp. $q/m \in (-r,0)$). This represents the strongest result up
to now available on the existence and multiplicity of connecting
solutions of (\ref{lorentz2}).

% and was indirectly
%obtained using a Kaluza-Klein framework.
%The results of \cite{piccione04} differ in
%the fact that they not only search for connecting solutions of
%(\ref{lorentz2}) but also for solutions satisfying the constraint
%(\ref{vin}), i.e. stationary points of $J$. We have seen that this
%condition means to discard all the solutions of Lorentz force
%equations that do not satisfy the physical (i.e. observable)
%constraint $\frac{q}{m} \int_{\eta} \dd s=\beta$. Of course, most
%interesting multiplicity results would be those related to the
%true Lorentz force equation (\ref{lorentz}), that is to the
%constraint (\ref{bla}).
\section{Other relations between LFE and EFE}
We point out in this section some other problems in which the
roles of the Lorentz force equation and the electromagnetic flow
equation could be confused.

\subsection{Symplectic formulation}
Equation (\ref{lorentz2}) is sometimes improperly referred as the
Lorentz force equation of a charge $Q$ in works that involve the
so called twisted symplectic form
\cite{guillemin84,sniatycki74,neitzke90}. Consider a spacetime $M$
having a metric $g_{\mu \nu}$ of signature $(+-\cdots-)$. On the
cotangent space $T^{*}M$ lives the canonical form $\Omega$ that in
local coordinates reads $\Omega=\dd p_{\mu} \wedge \dd q^{\mu}$.
Let $\pi: T^{*}M \to M$ be the canonical projection. On $T^{*}M$
we can define the twisted symplectic form $\Omega_{F}=\Omega+Q
\pi^{*} F$ where $F$ is the electromagnetic two-form. Let the
relativistic invariant (super)Hamiltonian be
$\mathcal{H}=\frac{1}{2}g^{\mu \nu}p_{\mu} p_{\nu}$. A
straightforward calculation shows that the integral lines of the
Hamiltonian flow, i.e. the integral lines of $X=\frac{\dd
x^{\mu}}{\dd \lambda} \frac{\p}{\p x^{\mu}}+\frac{\dd p_{\mu}}{\dd
\lambda} \frac{\p}{\p p_{\mu}}$ such that $i_{X} \Omega_{F}=-\dd
\mathcal{H}$, projected on $M$ are solutions of (\ref{lorentz2}).
This approach is surely fascinating in fact, contrary for instance
to the variational methods, here there is no reference to the
potential 1-form. However, the equation deduced is not the Lorentz
force equation so the same criticisms can be repeated here. The
true Lorentz force equation can be obtained from the Hamilton
equations using as Hamiltonian the relativistic energy.

\subsection{Jacobi fields}
Attention should also be paid on the different results available
for the study of the Jacobi fields for the two equations. Indeed
both \cite{caponio04c} and \cite{balakin00} present a calculation
of the deviation equation for the Lorentz force equation. However,
the two expressions differ since in the former case the
derivatives are with respect to a generic parameter while in the
latter case they are with respect to proper time. Indeed, in
\cite{caponio04c} what was actually calculated is the deviation
equation  for the {\em electromagnetic flow equation}. Since
solutions of the Lorentz force equation, even with different
charge-to-mass ratios, are solutions of the electromagnetic flow
equation, the Jacobi fields for the Lorentz force equation are
Jacobi fields for the electromagnetic flow equation while the
converse is not true. Moreover, in principle, there could exist a
Jacobi field of the electromagnetic flow equation which is not a
Jacobi field of the Lorentz force equation for no values of the
parameter $q/m$. This essentially because the Jacobi field may
actually be, in the space of solutions of the electromagnetic flow
equation, a tangent vector that connects solutions with different
values of the charge-to-mass ratio. Whether this possibility could
be indeed realized in some cases could be the  subject of further
investigation.

\subsection{The non-relativistic case}
We end the section with a digression on the non-relativistic
Lorentz force equation. Let us begin in an Euclidean space $E$ and
a spacetime $\Pi=E \times \mathbb{R}$ of coordinates $\{x^{i},t\}$
. It has the form
\begin{equation}
\frac{\dd^{2}x^{i}(t)}{\dd t^{2}}=\frac{q}{m} (
E^{i}+\epsilon_{ijk} \frac{\dd x^{j}(t)}{\dd t} B^{k}),
\end{equation}
which differ from the relativistic Lorentz force equation in
Minkowski spacetime only for a factor $1/\sqrt{1- (\dd x/\dd
t)^{2}}$ lacking at the left- hand side, between the two
derivatives. Now suppose that the electric field vanishes, then in
terms of the relativistic electromagnetic tensor the previous
equation reads
\begin{equation}
\frac{\dd^{2}x^{i}(t)}{\dd t^{2}}=\frac{q}{m} \hat F^{i}_{\ j}
\frac{\dd x^{j}}{\dd t},
\end{equation}
which admits a natural generalization in a curved {\em space} $S$
\begin{equation} \label{space}
D_{t} \frac{\dd x^{i}(t)}{\dd t}=\frac{q}{m} \hat F^{i}_{\ j}
\frac{\dd x^{j}}{\dd t},
\end{equation}
where $D$ is the covariant derivative on space compatible with the
space metric. This equation can be used to determine a flow in
$T^{*}S$ called {\em magnetic flow} \cite{burns02}, the projection
of trajectories in the flow being solutions of Eq. (\ref{space}).
Thus in this non-relativistic limit the {\em magnetic flow
equation} (\ref{space}) and the Lorentz force equation coincide.

Note that although Eq. (\ref{space}) is formally equivalent to
(\ref{lorentz2}) (indeed there is no constraint on the square of
${\dd x^{i}}/{\dd t}$ and $t$ may be regarded as an external
parameter) in a relativistic context the {\em electromagnetic flow
equation} and the Lorentz force equation differ. For this reason,
while it is quite natural to consider the problem of connectedness
of space points at a fixed times (that is the problem of
connecting two spacetime events) in the non-relativistic purely
magnetic limit, i.e. to look for parametrized solution of
(\ref{space}) that satisfy a constraint $x^{i}(t_{0})=x^{i}_{0}$
and $x^{i}(t_{1})=x^{i}_{1}$, the same formal problem for  Eq.
(\ref{lorentz2}) is less interesting since it has a completely
different interpretation (see the next section).

\section{A variational problem}
Consider the functional
\begin{equation} \label{j}
J_{x_{0},x_{1}}[\gamma]=\int _{\lambda_0}^{\lambda_1}
\left(\frac{1}{2} g(\gamma'(\lambda),\gamma'(\lambda)) + Q
\omega\left[\gamma'(\lambda)\right]\right)\dd \lambda.
\end{equation}
on the space of all the (absolutely continuous) causal curves,
which connect $x_0$ and $x_1$ in the interval
$[\lambda_0,\lambda_1]$. It generalizes the ``energy" functional
of Lorentzian geometry \cite{beem96,hawking73} to include a
vectorial potential. The energy functional contrary to the length
functional is well defined even for connecting curves whose causal
character changes with the parametrization. The connectedness of
spacetime through energy extremals has been studied deeply in the
mathematical literature providing an application of Morse and
Ljusternik-Schnirelman theory (see the survey \cite{sanchez01}).
The problem was then generalized to include a vectoral potential
as in (\ref{j}) (the works in section C of the bibliography are
related to this kind of problems). From the physical point of view
the fact that the functional $J_{x_{0},x_{1}}$ is defined
independently of the causal character of the curve makes it
difficult to establish the causal character of the extremals
although it enlarges its domain of applicability. Unfortunately,
it has been often claimed that if the extremal is timelike than it
is a solution of the LFE or equivalently that the functionals $J$
and $I$ are equivalent. However, a timelike stationary point
$\eta(\lambda)$ of $J_{x_{0},x_{1}}$ satisfies Eq.
(\ref{lorentz2}) and the constraint
\begin{equation} \label{vin}
x(\lambda_{0})=x_{0}, \quad x(\lambda_{1})=x_{1}.
\end{equation}
Let $\Delta \lambda=\lambda_{1}-\lambda_{0}$, and $\dd s/\dd
\lambda=C$, then integrating $C=(\int_{\eta} \dd s)/\Delta
\lambda$, thus the stationary point $\eta$ is a solution of the
Lorentz force equation (\ref{lorentz}) with charge-to-mass ratio
$q/m$ that satisfies
\begin{equation} \label{neo}
 \frac{q}{m}\int_{\eta} \dd s=Q \Delta
\lambda.
\end{equation}
Here $Q$ and $\Delta \lambda$ are fixed in the variational
principle but the length of the extremal is not fixed a priori and
therefore the charge-to-mass ratio of the extremal is not
determined a priori: different extremals will have different
charge-to-mass ratios. This happens because in order to fix the
charge-to-mass ratio one needs the constraint (\ref{bla}) while
the variational principle (\ref{j}) imposes the condition
(\ref{vin}) which implies that all the extremals have the same
product between charge-to-mass ratio and length. Thus, whatever is
the choice of the product $Q \Delta \lambda$, the existence of
stationary points of the action (\ref{j}) does not imply the
existence of a connecting solution of the Lorentz force equation
having a prescribed charge-to-mass ratio $q/m$. Using this
approach it is for instance not possible to prove the existence of
connecting trajectories for a charge-to-mass ratio like the one of
the electron or the one of the proton. Suppose one proves that
(\ref{j}) admits a timelike extremal: it can actually have a
charge-to-mass ratio that does not corresponds to an existing
particle. More generally, the same happens if one proves that Eq.
(\ref{lorentz2}) has a connecting solution. For this reason the
physical interpretation of Eq. (\ref{lorentz}) and Eq.
(\ref{lorentz2}), and the variational principles (\ref{fun}) and
(\ref{j}) is different and in general to have a strict contact to
physical questions (\ref{lorentz}) or (\ref{fun}) should be used.
As another example suppose we wish to study in how many ways an
electron can leave an event  $x_{0}$ to reach an event $x_{1}$,
then we should clearly study how many extremals the
charged-particle action has. On the contrary if one proves that
the action (\ref{j}) has say, four extremals, it could be that
none of them has the charge-to-mass ratio of the electron.

Let us consider in more detail the action (\ref{j}). Note that the
extremals when regarded as unparametrized curves depend only on
the product $Q\Delta \lambda$ of (\ref{j}). In other words given
$\beta=Q\Delta \lambda$ two choices of the action (\ref{j}) with
the same $\beta$ have the same extremals up to reparametrizations.
 We said that its timelike extremals, when regarded as
trajectories, are solutions of the Lorentz force equation
(\ref{lorentz}) for a charge-to-mass ratio that satisfies the
constraint $\frac{q}{m} \int_{\eta} \dd s=\beta$. Conversely,
given a timelike connecting solution of (\ref{lorentz}), $\eta$,
that satisfies the constraint (\ref{neo}) a parametrization can be
found so that $\eta(\lambda)$ becomes a timelike extremal of
(\ref{j}): it suffices to choose the parametrization such that
$\dd \lambda=\frac{q}{m Q}\dd s $ and $\eta(\lambda_{0})=x_{0}$.

We give now a variational principle that has the same
unparametrized extremals of (\ref{j}). Consider the functional
\begin{equation} \label{k}
K_{x_{0}, x_{1}}[\gamma]=\frac{1}{2}\left(\int _{\gamma} \dd s
\right)^{2}+\beta \int _{\gamma} \omega
\end{equation}
defined on the set of $C^{1}$ causal curves connecting $x_{0}$ and
$x_{1}$. A computation of the Euler-Lagrange equation immediately
shows that the timelike  extremals of this functional are those
timelike curves which satisfy the Lorentz force equation
(\ref{lorentz}) having a charge-to-mass ratio and a length that
satisfy the constraint $\frac{q}{m} \int_{\eta} \dd s=\beta$, i.e.
they are the same, but this time unparametrized, extremals of
(\ref{j}). This functional removing the unobservable dependence on
the parametrization could help to reveal more clearly the physical
meaning of (\ref{j}). For fixed $q/m$ and $\beta$ (\ref{fun}) and
(\ref{k}) (that is (\ref{j})) have in general different extremals,
however let $\eta$ be an extremal of the charged-particle action,
and { \em choose} $\beta=\frac{q}{m} \int_{\eta} \dd s$ then
(\ref{k}) and therefore (\ref{j}) will have $\eta$ as extremal. In
other words, for each connecting solution $\eta$ of the Lorentz
force equation of charge-to-mass ratio $q/m$ there is a choice of
$\beta$ such that $\eta$ is an extremal of $K$ (or, which is the
same, $J$) for that $\beta$. Thus no connecting solution of the
Lorentz force equation is left out considering the extremals of
$K$ for all the values of $\beta \in \mathbb{R}$. The problem is
that they are classified according to a parameter $\beta$ which is
not as interesting as the charge-to-mass ratio $q/m$ is. It is
interesting to note that since $ 0 \le \int_{\eta} \dd s \le
l(x_{0},x_{1})$, where $l(x_{0},x_{1})$ is the Lorentzian distance
function, Eq. (\ref{neo}) implies that
\begin{equation}
\vert \frac{q}{m} \vert \ge \frac{\vert
\beta\vert}{l(x_{0},x_{1})}
\end{equation}
that is, the variational principles (\ref{j}), (\ref{k}), for a
given $\beta$ have  as timelike stationary points solutions of
Lorentz force equations with charge-to-mass ratios having an
absolute value bounded from below.

\subsection{Erratas and other comments} It seems that some
confusion regarding the use of the Lorentz force equation and its
interpretation started from the work \cite{benci98} where the
authors introduced the functional (\ref{j}). In this respect it is
better to point out some erratas that may lead to improper
interpretations. They show that an extremal point $x(\lambda)$ of
the action (the same as $J$ but without the factor $1/2$)
\begin{equation} \label{temp}
\tilde{J}_{x_{0},x_{1}}[\gamma]=\int _{\lambda_0}^{\lambda_1} \{
g(\gamma'(\lambda),\gamma'(\lambda)) + Q
\omega\left[\gamma'(\lambda)\right] \}\dd \lambda
\end{equation}
has a constant square of the 4-velocity that they call $m$ as in
(\ref{bla}). A proof that $x(\lambda)$ is also an extremal point
of (\ref{fun}) with $q/m=Q/m$ was also claimed, but unfortunately
this statement is true only if $\tilde{J}$ is replaced with $J$.
In fact, (we use our notation) knowing that the stationary point
satisfies (\ref{bla}) for a certain $m$, they use this in
$\tilde{J}$ to rewrite
\begin{equation}
 g(\gamma'(\lambda),\gamma'(\lambda))=(\frac{\dd s}{\dd
 \lambda})^{2} =m \frac{\dd s}{\dd
 \lambda}
\end{equation}
and replacing in (\ref{temp}) obtain $mI_{x_{0},x_{1}}$. Then they
go on to calculate the Euler-Lagrange equation of (\ref{fun})
assuming that this should be equivalent to the initial one.
However, it is well known that this way of working is incorrect
and in fact the two variational principles $\tilde{J}$ and $I$ so
constructed, do not necessarily share an extremal point. Indeed,
using $\dd s= m \dd \lambda$ and the Euler-Lagrange equation for
$\tilde{J}$ we find that the trajectory $x$ satisfies the Lorentz
force equation with charge-to-mass ratio $Q/(2m)$ while in order
to be an extremal of the obtained $I$ it should satisfy it with
charge-to-mass ratio $Q/m$. In general it is incorrect to replace
inside the variational principle information that follows from its
Euler-Lagrange equation as the new variational principle so
obtained does not have the same stationary points.

This work generated some confusion in subsequent literature. For
instance  in \cite{caponio02b} the authors include the $1/2$
factor but then they state \cite[Remark 1.1]{caponio02b} (see also
\cite[Remark 1.2]{caponio02} and \cite[p. 128]{bartolo02}) that
the functionals $I$ with $q/m=1$ and $J$ with $\beta=1$ have the
same stationary points up to reparametrizations. They refer for a
proof to \cite{benci98}. However, this statement is incorrect
since, as we said above, it is true that each extremal of $J$,
with $\beta=1$, is extremal of $I$ for a certain, unknown a
priori, $q/m$, and it is true that an extremal of $I$, with
$q/m=1$, is an extremal of $J$ for a certain, unknown a priori,
$\beta$, but this does not imply that $J$ with $\beta=1$ and $I$
with $q/m=1$ have the same stationary points. The corrected proof
of the modified statement was given in \cite{antonacci00}. We
stress that in any case this problem did not affect the
mathematical conclusions of those works although it severely
restricts the physical implications.

\section{Existence results and conclusions}
Let us come to the existence results available.
%We point out that
%thanks to the previous analysis their physical meaning will be
%clearly distinguished while in the original articles where those
%results appeared they were simply referred generically as results
%on ``the Lorentz force equation".
%On the contrary, as we have
%shown, we are sometimes in presence of an equation distinct from
%(\ref{lorentz}), and in case of a variational approach, in
%presence of constraints different from (\ref{bla}); the physical
%interpretation then changes accordingly.
A first result was obtained for the existence of connecting
solutions of (\ref{lorentz2}). In \cite{caponio01,caponio04d} it
was proved that (\ref{lorentz2}) has always a connecting solution
in a globally hyperbolic spacetime. An  analogous result for Eq.
(\ref{lorentz}) was given in  \cite{minguzzi03b} (first relevant
advances in \cite{caponio03}). This implied in particular the
existence of a maximum for the charged-particle action (\ref{fun})
and could be read as a multiplicity result for Eq.
(\ref{lorentz2}).

The work on the action $J$ began in \cite{bartolo99,antonacci00},
the action $J$ being a natural generalization of the ``energy"
functional  of Lorentzian geometry to include a vectorial
potential. This allows one to consider geometrical questions that
otherwise could not be implemented using $I$, for instance the
spacetime connectedness through {\em spacelike} extremals of $J$.
From the physical point of view, however, most interesting are
timelike extremals and in this respect existence results for $J$
are up to now weaker than those for $I$ (in a globally hyperbolic
spacetime, for instance, as far as we know there could be no
timelike extremals for certain values of $\beta$), although
related results have been obtained for stationary spacetimes
\cite{bartolo00,bartolo01,caponio02,bartolo02,bartolo03,mirenghi02},
time periodic potentials and metrics \cite{caponio04},  or under
other assumptions \cite{bartolo99,antonacci00}.

Although action $J$ has a good behavior under standard variational
methods and Morse theory, and it gives rise to some interesting
mathematical problems, we believe that it should not be studied as
a substitute for $I$. Indeed, the variational difficulties for $I$
are now circumvented using a geometrical interpretation
\cite{minguzzi03b} that makes it possible to use causal
techniques. Moreover, even if the results for $J$ and $I$ were
comparable, the physical interpretation of $J$'s timelike
extremals, that we previously pointed out, would not allow to make
contact with realistic charge-to-mass ratios.

Finally, in order to clarify the relation between different
articles we consider three existence problems for trajectories
connecting the events $x_{0}$ and $x_{1}$. We regard each one as
the problem of finding a connecting solution of
\begin{itemize}
\item[A.] The electromagnetic flow equation (\ref{ahh}) (or, which is the same (\ref{lorentz2})).
\item[B.] The Lorentz force equation (\ref{lorentz}).
\item[C.] The equation (\ref{lorentz2}) with the constraint (\ref{vin}).
\end{itemize}

\begin{widetext}
\begin{center}
\begin{table}
    \begin{center}
        \begin{tabular}{|c|c|c|c|c|c|}
            \hline
            & & & & &\\
             {\bf Problem}  &  {\bf Equations} &   {\bf Parameters} &  {\bf Functional} &  {\bf Physical constraint} &  {\bf Literature}\\
              & & & & &\\
            \hline\hline
            A. & (\ref{ahh}) &  $\epsilon=\pm 1$ & - & $\frac{\textrm{q}}{\textrm{m}}(\eta)=R \quad \textrm{or} \quad
            \textrm{sgn}(\frac{\textrm{q}}{\textrm{m}}(\eta))=\epsilon$  &  \cite{caponio01}-\cite{masiello02} \\
            \hline
            B. & (\ref{lorentz}) & $q/m \in \mathbb{R}$ & $I$  & $\frac{\textrm{q}}{\textrm{m}}(\eta)=R \quad \textrm{or} \ \ \qquad \frac{\textrm{q}}{\textrm{m}}(\eta)=\frac{q}{m}$  &  \cite{caponio03}-\cite{minguzzi03b} \\
            \hline
            C. & (\ref{lorentz2}) and (\ref{vin}) & $\beta \in \mathbb{R}$ & $J$ or $K$  & $\frac{\textrm{q}}{\textrm{m}}(\eta)=R \quad \textrm{or} \quad
\frac{\textrm{q}}{\textrm{m}}(\eta)\int_{\eta} \dd s=\beta$ &
            \cite{antonacci00}-\cite{mirenghi04}\\
%& & & & &
%\cite{caponio02,caponio02b,caponio04,caponio04c,mirenghi02,mirenghi04}
%\\
            \hline
        \end{tabular}
    \end{center}
    \caption{Different existence problems for the connecting solutions. Case B is the one of the Lorentz force equation. Let $\beta$ and $q/m$ be given.
    A solution of B is a solution of A but not necessarily of C. A solution of C is a solution of A but not necessarily of B. A solution of A is not necessarily  a solution of B or C.}
\label{tab}
\end{table}
\end{center}
\end{widetext}

Table \ref{tab} presents the three different existence problems
pointing out if they have a variational Lagrangian formulation, on
which parameters the functional depends, what is the physical
constraint on  the charge-to-mass ratio and what are the works
that dealt or that are related with that problem.

In conclusion we believe to have clarified the  mathematical  and
physical aspects of different problems  considered in the
literature. Although each one has  something related to the
Lorentz force equation, attention should be paid since the results
available have different mathematical and physical meanings.

\begin{acknowledgments}
I would like to thank E. Prati and M. S\'anchez for useful suggestions. The author is supported by INFN, grant $\textrm{n}^{\circ}$ 9503/02.\\
\end{acknowledgments}

%\newpage

%\bibliography{../../bibliografie/simultaneity,../../bibliografie/libri} \bibliographystyle{plain}

\begin{thebibliography}{10}
\bibitem{endnote30}{Of course, only as long as the physics of the
Lorentz force equation is concerned. At the variational level one
should add more terms to the charged-particle action in order to
make both charge and mass observables. This, however, would lead
us to a different mathematical problem.}
\end{thebibliography}

\end{document}